\documentclass{article}

\usepackage{arxiv}

\usepackage[utf8]{inputenc} 
\usepackage[T1]{fontenc}    
\usepackage{hyperref}       
\usepackage{url}            
\usepackage{booktabs}       
\usepackage{amsfonts}       
\usepackage{nicefrac}       
\usepackage{microtype}      

\usepackage{lipsum}         
\usepackage{graphicx}  
\usepackage[numbers,sort&compress]{natbib}
\usepackage{doi}
\usepackage{amsmath}
\usepackage{cleveref}       

\title{Early Stages of Self-Healing at Tungsten Grain Boundaries from Ab Initio Machine Learning Simulations}

\author{
  Jorge Suárez-Recio \\
  Instituto de Fusión Nuclear “Guillermo Velarde”, C. de José Gutiérrez Abascal 2, 28006 Madrid, Spain \\
  Departamento de Ingeniería Energética, Universidad Politécnica de Madrid, 28006 Madrid, Spain \\
  Universidad de Oviedo, C. Leopoldo Calvo Sotelo 18, 33007 Oviedo, Spain \\
  \texttt{j.srecio@upm.es}
  \And
  Pablo M. Piaggi \\
  CIC nanoGUNE BRTA, Tolosa Hiribidea 76, 20018 Donostia-San Sebastián, Spain \\
  Ikerbasque, Basque Foundation for Science, 48013 Bilbao, Spain \\
  \And
  Francisco J. Domínguez-Gutiérrez \\ 
  National Center for Nuclear Research, ul. A. Soltana 7, 05-400 Otwock, Poland \\
  \And
  Raquel Gonzalez-Arrabal \\
  Instituto de Fusión Nuclear “Guillermo Velarde”, C. de José Gutiérrez Abascal 2, 28006 Madrid, Spain \\
  Departamento de Ingeniería Energética, Universidad Politécnica de Madrid, 28006 Madrid, Spain  \\
  \And
  Roberto Iglesias \\
  Universidad de Oviedo, C. Leopoldo Calvo Sotelo 18, 33007 Oviedo, Spain \\
  Centro Universitario ASturias RAw Materials Institute (ASRAM), C. Gonzalo Gutiérrez Quirós, 33600 Mieres, Spain  
}

\begin{document}
\maketitle

\begin{abstract}
	Nanostructured tungsten has been reported as a possible alternative plasma‐facing material due to its potential ability to self‐heal radiation-induced defects, a property that is attributed to its high density of grain boundaries (GB). Here, we study the initial stages of self-healing at tungsten interfaces with molecular dynamics simulations driven by a machine-learning interatomic potential tailored to one of the most common GBs found in experiments. Our model accurately reproduces the \textit{ab initio} potential energy surface derived from density functional theory (DFT) calculations and outperforms previously reported empirical and machine learning interatomic potentials in predicting defect energetics. The simulations reveal low-temperature defect migration to GBs driven by rapid dumbbell-like ordering and subsequent accommodation along GB grooves. In contrast to empirical potentials, which predict unexpected GB degradation at high temperatures after defect migration, our model maintains stable GB motifs over the investigated temperature range. The temperature-dependent defect counts, evaluated using an Arrhenius-like fit, yield an average interstitial migration energy of 0.048 eV, in agreement with experiment.

This work underscores the capabilities of \textit{ab initio} machine learning simulations in accurately modeling defect-GB interactions and highlights their potential to contribute to the development of radiation tolerant materials.
\end{abstract}

\keywords{tungsten \and self healing \and molecular dynamics \and machine learning interatomic potential \and nanostructured materials}

\section{Introduction}

Decarbonizing energy systems has become a defining challenge of the 21st century. Although fission reactors currently supply about 10\% of the world's electricity \cite{IEA_Nuclear} and continue to evolve with advanced designs, fusion energy is poised to address the interconnected crises of climate change, energy security, and resource competition. Its high power density, lack of long-lived radioactive waste, and inherent safety (fusion reactions stop naturally under abnormal conditions, eliminating the risk of reactor runaway) make fusion a promising route to next-generation power plants. Realizing this promise on a commercial scale, however, requires the development of robust plasma-facing materials capable of withstanding intense radiation and extreme thermal loads, hence acting as the reactor's first line of defense. Ideally, they should exhibit healing properties that allow radiation-induced defects to be eliminated, thereby preserving structural and functional integrity. Progress toward such materials requires not only breakthroughs in fabrication and engineering, but also the development of computational methods to predict their behavior under high temperature and high-flux conditions, harsh environments that are still difficult to fully replicate in existing experimental facilities.

Irradiation of coarse-grained materials by particles (e.g., ions, neutrons) with energies exceeding the displacement threshold, as occurs in the inertial nuclear reactors operated in the direct drive approach, produces atomic displacements that generate vacancies and self-interstitial atoms (SIAs). At a given temperature, both types of defects migrate towards defect sinks such as grain boundaries (GBs), dislocations, and voids. However, in general, they do so at quite different energies and time scales. These differences mean that they usually do not meet and recombine, leading to a permanent degradation of material properties such as electrical conductivity and mechanical response \cite{Chalmers1949,Coble1963,ASHBY1969837,GLEITER1979187,Okamoto19792,WIEDERSICH197998,Lane01121983,YAMAKOV200261,KIRCHHEIM2002413,Demkowicz2012,doi:10.1126/science.1224737}. From an atomistic point of view, the situation can be different if vacancies and SIAs are forced to be very close to each other, which can be done by reducing the grain size (nanostructuring) or, in other words, by introducing a high density of GBs. In this case, the recombination probability between vacancies and SIAs increases, leading, under certain conditions, to their mutual annihilation at the surface or GBs, a phenomenon commonly referred to as self-healing, so that the material spontaneously returns to its unirradiated structure. Nevertheless, as previously mentioned, this self-healing behavior only occurs under specific conditions, well captured phenomenologically by the three-temperature (low, medium, and high) model proposed by Beyerlein \textit{et al.} \cite{BEYERLEIN2013443}. In the low temperature regime, vacancies are immobile, so any damage inflicted on the nanostructured material is greater than on its coarse-grained counterpart \cite{doi:10.1126/science.1183723,Bai2013}. This is because highly mobile SIAs migrate along the bulk toward the GBs, leaving immobile vacancies in the grain interior. At intermediate temperatures, where vacancies are still immobile, some of the radiation-induced damage can be annihilated by the re-emission of SIAs from the GBs \cite{Bai2013,PhysRevB.85.012103,Uberuaga_2013}. Finally, at high temperatures, vacancies become mobile and move toward GBs, where they effectively recombine with SIAs, annihilating the radiation-induced damage. The evolution of defects and the temperature threshold for moving from one temperature regime to another depend on the material composition and microstructure.

Even though this model describes the origin of the self-healing behavior, its effectiveness depends strongly on the GB architecture \cite{558899744e284bca9c552cef31fd5b40}. Therefore, a proper description of the self-healing capabilities of nanostructured materials requires an understanding of how defects migrate and interact with the specific GB at the atomic scale, which can be achieved by performing computer simulations.

Traditionally, two main approaches have been used: \textit{ab initio} molecular dynamics (AIMD) — typically based on Density Functional Theory (DFT) to compute interatomic forces — and classical Molecular Dynamics (MD) using empirical or semi-empirical interatomic potentials (hereafter IPs). While AIMD offers high precision in capturing atomic interactions, it is constrained by limited time and length scales \cite{Allen2017,Marx2009}.  Conversely, classical MD allows the simulation of larger systems over longer timescales, but relies on IPs that often have limited accuracy and transferability \cite{GRIGOREV2016143,PhysRevMaterials.7.043603}. As a result, an IP developed for coarse-grained materials may not be sufficiently accurate to adequately describe radiation-induced damage in nanostructured materials. Indeed, this fact could limit self-healing studies, where the simulation of defect migration requires an extremely precise description of GB energetics as well as relatively large simulation boxes and long simulation times. 

Recent years have witnessed the rapid development of machine learning interatomic potentials (MLIPs) that leverage large datasets, often derived from quantum-mechanical calculations, to build flexible, highly accurate models of atomic interactions \cite{Bartok2010,Thompson2015SNAP,Drautz2019}.
These advanced methods have already been applied to challenging problems such as radiation damage cascades, surface reactions, and phase transformations under extreme thermomechanical conditions \cite{Deringer2019MLIPs,Novikov2021MLIPs,PhysRevB.85.214103,gong2024ml_nickel}. Critically, they can achieve near-\emph{ab initio} accuracy at a fraction of the computational cost, thus enabling simulations on time and length scales relevant to capturing the early stages of self-healing at tungsten interfaces.

In this work, we present a computational study of how intrinsic defects migrate and interact with a tungsten GB, with an eye towards having a deeper understanding of the self-healing in nanostructured tungsten, one of the candidates proposed as a plasma-facing material in nuclear fusion reactors \cite{marchhart2024discovering,wu2019manufacturing,ELATWANI2018206,Fukuda2013,janeschitz2001plasma}. We focus on atomistic damage since the thermomechanical effects that PFMs will undergo and that could lead to significant temperature increases and, consequently, grain growth (if this is not somehow stabilized), could be alleviated by using a/or by combining different strategies: lower target yield, employing radiation mitigation strategies, and optimizing the chamber geometry. However, atomistic effects cannot be avoided. To study the latter, we develop a IP, dubbed MLIP-W, based on the Deep Potential methodology \cite{Zhang2018} and trained on quantum mechanical DFT data.
Although some MLIPs for tungsten have been developed in recent years \cite{wood2017quantumaccuratemoleculardynamicspotential,Sikorski2023Machine,Byggmastar2019Machine}, none of them have been specifically trained to describe GBs meaning that they may not be guaranteed to preserve \textit{ab initio} accuracy for complex environments involving both defects and interfaces. 
We apply our model to study the early stages of self-healing wherein SIAs migrate to GBs. Because the nanosecond-scale MD simulations carried out here are orders of magnitude shorter than the second-scale vacancy diffusion times, vacancy motion cannot be resolved and is not considered in our study. We explicitly train our potential to accurately describe the W\(\langle 110 \rangle\)/W\(\langle 112 \rangle\) GB over a wide temperature range, as well as the interaction of this GB with point defects. It is worth mentioning that we select this particular GB because it is the most frequently observed in sputter-deposited tungsten films, a fabrication route that can be easily scaled up for industrial production \cite{Gordillo2014}. We then compare the performance of MLIP-W to two recent, state-of-the-art MLIPs, as well as to widely used empirical potentials, to highlight its advantages. Moreover, we perform automated defect identification and classification using a fingerprinting toolkit that considers the local atomic environments \cite{VONTOUSSAINT2021107816,Bartok2013}.
Our results show that the MLIP-W outperforms existing IPs in capturing defect migration. Moreover, the MLIP-W also preserves the structure of the GB at high temperatures, in contrast with previous IPs. These findings show the effectiveness of the MLIP-W in describing self-healing near GBs, offering a more accurate computational framework for modeling radiation damage.

\section{Results}

\subsection{Development of the MLIP-W}
\label{subsec:development_deepmd}

To accurately capture the \textit{ab initio} potential energy surface (PES) for W, we used the smooth version of the Deep Potential method \cite{Zhang2018,Zhang18end} as implemented in DeePMD-kit \cite{WANG2018178,10.1063/5.0155600}, an open-source deep learning software for training neural network potentials. In our workflow, we coupled this methodology with an active learning scheme to iteratively refine the training set and generate an extensive set of atomic configurations and their corresponding DFT-derived energies and forces \cite{Zhang_2019,Smith2018JCP}. This methodology has been successfully demonstrated on several occasions, yielding excellent results \cite{Piaggi2024Faraday,Schran2021PNAS}. At each step of the active learning process, short MD simulations were run to identify "high-error" configurations based on force discrepancies among a set of independently trained networks. These uncertain configurations were recalculated using DFT, and the model was retrained accordingly. A detailed explanation of the active learning approach is provided in Subsection~\ref{subsec:Training_DeePMD}.

\begin{figure}[t!]
\centering
\includegraphics[width=1\linewidth]{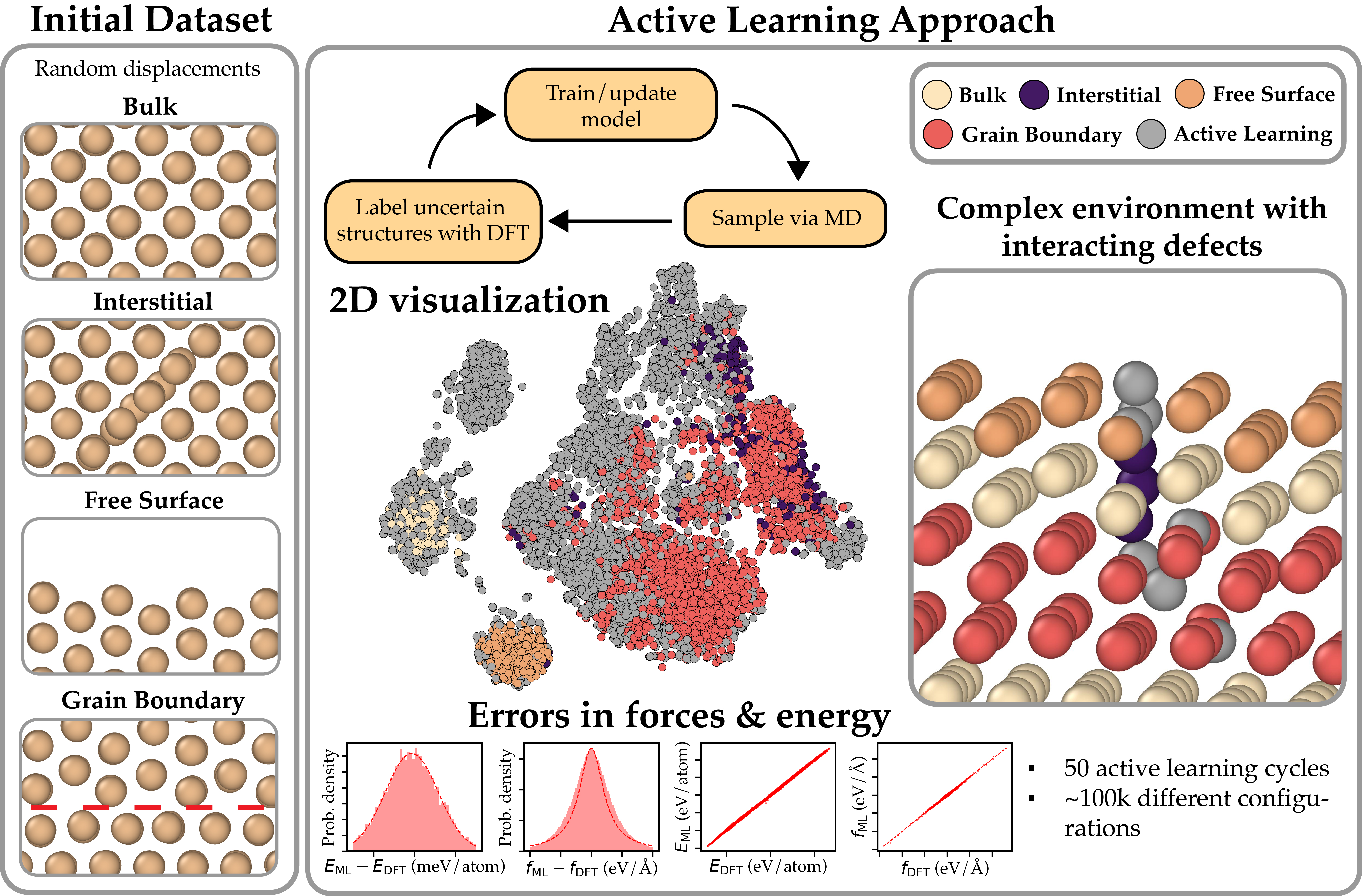}
\caption{\textbf{Development of the MLIP-W.} The left panels depict representative atomic environments included in the initial training dataset, which is composed of configurations generated by random displacements and single point defects. At the top of the center panel, a simplified flowchart outlines the active-learning loop used to enrich the dataset.
Below that, a t-SNE visualization of local atomic environments encoded by DeePMD descriptors reveals the following: bulk-like environments (beige), interstitial defects (purple), free surface atoms (orange), GB atoms (red), and environments sampled via active learning (gray). The panel on the right provides an example of the complexity and diversity of the atomic environments in the dataset by showing the presence of multiple defect types in a single configuration. The error distributions and parity plots at the bottom quantify the overall accuracy achieved after approximately 50 active learning cycles.}
\label{training_data}
\end{figure}

In our case, the initial training dataset includes bulk W configurations and the W$\langle110\rangle$/W$\langle112\rangle$ GB, with random atomic displacements applied to sample regions of the PES near equilibrium positions (see the left panels of Fig.~\ref{training_data}). To accurately capture intrinsic defect behavior, we randomly introduced vacancies and SIAs into both the bulk and GB configurations. However, to further explore GB-defect interactions at finite temperatures, we followed an active learning approach to identify and include new atomic environments critical to our study. The central panel of Fig.~\ref{training_data} presents a t-distributed stochastic neighbor embedding (t-SNE) visualization of these environments. This visualization was generated from a subset of 10,000 descriptor vectors (DVs)—which are numerical representations of local atomic environments—drawn from the full training set. The environments identified and integrated via active learning are shown in gray. They typically correspond to highly complex scenarios involving multiple defect interactions, such as interstitial atoms interacting with GBs or free surfaces, as shown by the dashed circles in the panel on the right. The integration of active learning with the flexible MLIP framework substantially improves upon the limitations of empirical potentials, which are characterized by a closed-form function, enabling the treatment of highly nontrivial defect-GB scenarios.

To ensure the comprehensiveness of our dataset and to minimize potential biases or gaps, we continued the active learning process iteratively, completing a total of 50 cycles, until no new configurations exhibiting uncertainties above our predefined threshold were identified in subsequent MD simulations. Moreover, convergence in the descriptor distribution was observed, with new configurations consistently falling inside the existing descriptor space. This indicates that no new significant atomic environments were discovered in the later cycles of active learning.

To measure its overall accuracy, we computed the root mean square (RMS) errors for energies and forces over the entire training set, obtaining $\epsilon_E^{\mathrm{RMS}} \approx 0.84$\,meV/atom and $\epsilon_F^{\mathrm{RMS}} \approx 151$\,meV/\AA. These errors compare favorably to other state-of-the-art MLIPs \cite{PhysRevLett.126.236001,ccc7602f6d754fa199a1835dfbc3bf26}. The bottom panels of Fig.~\ref{training_data} include the error distributions for energies and forces on the left, followed by parity plots comparing the ML- and DFT-predicted energies and forces, respectively, on the right. This performance highlights the ability of the model to 
capture subtle features of the PES, especially in systems where 
the geometry deviates significantly from the bulk, such as 
defects, GBs, and surfaces. 
Note that the reference DFT calculations were performed using the PBE exchange-correlation functional. Therefore, the accuracy of the derived potential remains inherently limited by this specific choice.

\subsection{Structural properties and defect energetics benchmarks}
\label{subsec:validation_deepmd}

Once the training of the MLIP-W was completed,
the next step was to systematically evaluate its performance to confirm the predictive capabilities and its ability to accurately reproduce the PES in diverse atomic environments. 

Table ~\ref{comparison_pot} compares several fundamental properties of BCC W as predicted by our MLIP-W model, DFT (computed in this work), experiments (where available), two widely used empirical IPs—an embedded atom method (EAM) potential by Marinica \textit{et al.}~\cite{marinica2013} and a Tersoff potential by Juslin \& Wirth~\cite{Juslin2013}, and by two recent MLIPs—the tabulated Gaussian Approximation Potential (TabGAP) by Byggmästar \textit{et al.}~\cite{Byggmastar2019Machine} and the Spectral Neighbor Analysis Potential (SNAP) by Wood \textit{et al.}~\cite{wood2017quantum}. Our model closely matches DFT results for all properties and outperforms other models. MLIP-W predicts the single vacancy formation energy, \(E_f(\mathrm{V})\), with a deviation of \(0.04\) eV with regards to the DFT values, and the \(\langle110\rangle\) and \(\langle111\rangle\) dumbbell formation energies, \(E_{f,db110}\) and \(E_{f,db111}\), with deviations of \(0.07\) eV and \(0.10\) eV, respectively. The octahedral interstitial formation energy, \(E_{f,octa}\), is underestimated by \(0.31\) eV relative to DFT. Although the last deviation is larger than the others, the value calculated with the MLIP-W remains the closest to the DFT reference. The EAM potential shows the next smallest discrepancy, of 0.38 eV. For the first-nearest-neighbor di-vacancy binding energy, \(E_b(V\!-\!V\ 1\mathrm{nn})\), the Tersoff (\(+0.06\) eV), EAM (\(+0.08\) eV) and SNAP (\(+0.07\) eV) potentials produce values closer to the DFT reference than MLIP-W (\(-0.12\) eV). However, MLIP-W is the only model, apart from SNAP, which gives a value of \(-0.07\) eV, to reproduce the negative second-nearest-neighbor di-vacancy binding energy, \(E_b(V\!-\!V\ 2\mathrm{nn})\), reporting \(-0.21\) eV versus \(-0.19\) eV in DFT. We also evaluated the generalized stacking fault energy (GSFE) profiles, which are critical for dislocation-mediated defect motion. Though MLIP-W was not trained on GSFE data, we found that it closely reproduces DFT barrier heights on both the \{110\} and \{112\} planes (see Supplementary Figures S1B–C) \cite{marinica2013,Juslin2013,Thompson2015,Bonny2014}. This agreement confirms the reliability of MLIP-W regarding the energetics that govern defect migration. For full details, see Supplementary Note 1.

\begin{table*}[b!]
\centering
\caption{Tungsten lattice parameter ($a_0$), formation energy for a single vacancy ($E_f$) and various interstitial configurations including dumbbells in the $\langle110\rangle$ ($E_{f,db110}$) and $\langle111\rangle$ ($E_{f,db111}$) directions, and in an octahedral site ($E_{f,octa}$), and di-vacancy binding energies in first nearest neighbors ($E_b$(V--V) 1nn) and second nearest neighbors ($E_b$(V--V) 2nn) arrangements. We compare results obtained experimentally and calculated using DFT, MLIP-W (this work), Tersoff \cite{Juslin2013}, EAM \cite{marinica2013}, TabGAP \cite{Byggmastar2019Machine}, and SNAP \cite{wood2017quantum}.}
\small
\begin{tabular}{c c c c c c c c}
\hline\hline
\textbf{Property} &
\textbf{Exp.} &
\textbf{DFT} &
\textbf{MLIP-W} &
\textbf{Tersoff} &
\textbf{EAM} &
\textbf{TabGAP} &
\textbf{SNAP} \\ \hline
$a_0$ (BCC) (\AA)      & 3.17 \cite{kittel1976introduction,lassner1999tungsten,lide2004handbook} & 3.17  & 3.16 & 3.16 & 3.14 & 3.18 & 3.18 \\[2pt]
$E_f$(V) (eV)          & 3.15 \cite{GORECKI1977} & $3.34$ & 3.30 & 3.71 & 3.49 & 3.22 & 3.20 \\[2pt]
$E_{f,db110}$ (eV)     &  & $11.04$ & 10.97 & 10.16 & 10.86 & 10.64 & 9.76 \\[2pt]
$E_{f,db111}$ (eV)     &  & $10.76$ & 10.86 & 9.50  & 10.40 & 10.34 & 9.55 \\[2pt]
$E_{f,octa}$ (eV)      &  & $13.11$ & 12.80 & 10.39 & 12.73 & 10.50 & 11.5 \\[2pt]
$E_b$(V--V) 1nn (eV)   & 0.7 \cite{Park01031983} & $0.41$ & 0.29 & 0.47 & 0.49 & 0.06 & 0.48 \\[2pt]
$E_b$(V--V) 2nn (eV)   &  & $-0.19$ & $-0.21$ & 0.41 & 0.38 & 0.05 & $-0.07$ \\ 
\hline\hline
\end{tabular}
\label{comparison_pot}
\end{table*}

\begin{figure}[t!]
\centering
\includegraphics[width=1.0\linewidth]{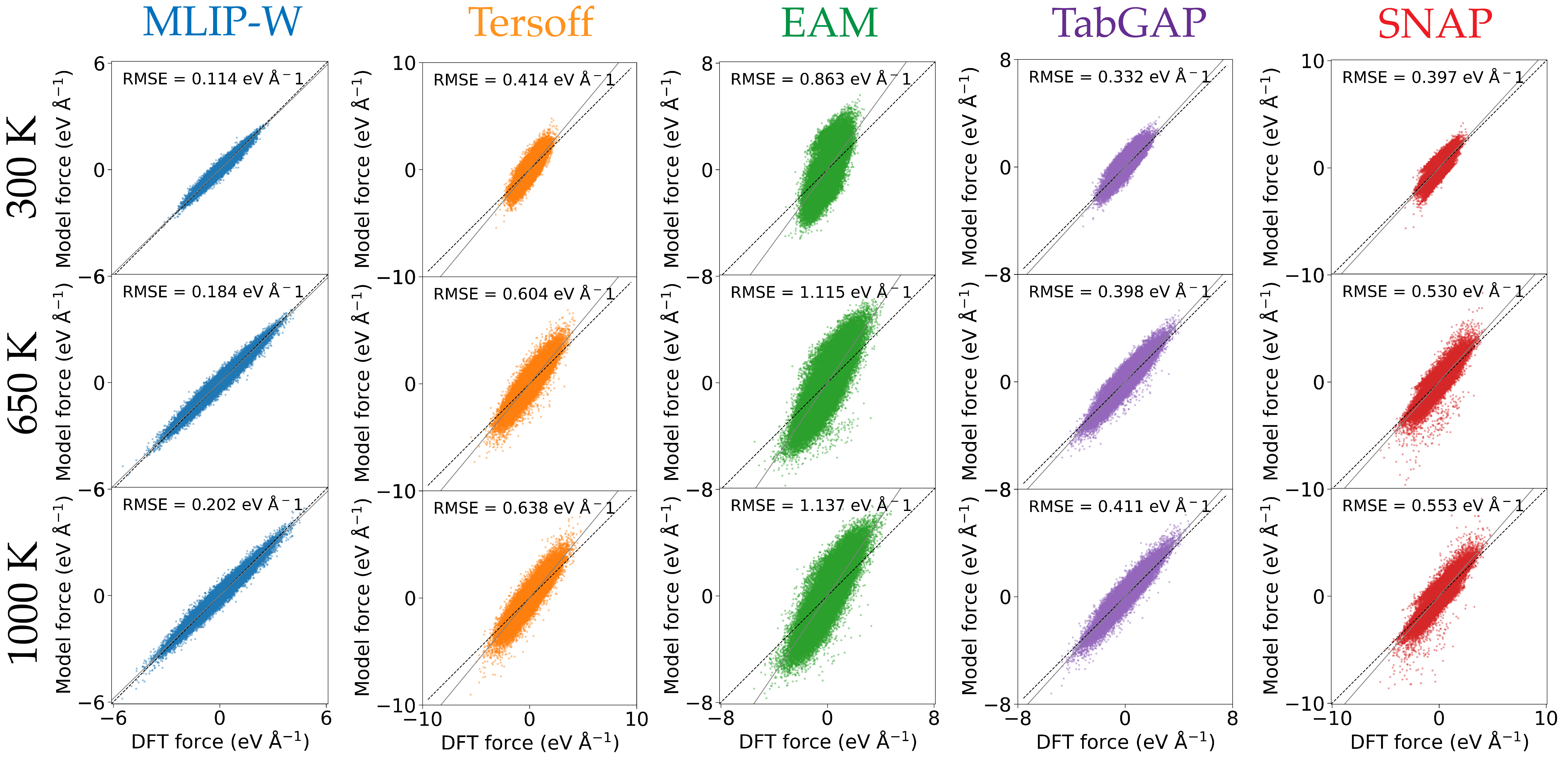}
\caption{\textbf{Finite-temperature force accuracy of the evaluated potentials.} Parity plots comparing the atomic forces predicted by each IP (vertical axis) with the reference DFT forces (horizontal axis). The plots are for W$\langle110\rangle$/W$\langle112\rangle$ GB configurations containing Frenkel pairs. The configurations were sampled after MD evolution at 300 K (top row), 650 K (middle row), and 1000 K (bottom row). The columns correspond to MLIP-W, Tersoff, EAM, TabGAP, and SNAP from left to right. The solid black line shows perfect
correlation, and each panel reports the RMSE.}

\label{parity_plot_new}
\end{figure}

However, while these static benchmarks provide an initial validation, and even parameterizations for larger time and length scale models, they alone do not necessarily guarantee accurate defect kinetics or correct long-term microstructural evolution. The next essential step is to evaluate the performance of the MLIP-W and the other potentials under finite-temperature conditions that are directly relevant to our simulations of defect-GB interactions and their dynamics. To this end, we took the W$\langle110\rangle$/W$\langle112\rangle$ GB described above, randomly introduced interstitials and vacancies, and evolved the system using our model at 300, 650, and 1000 K to obtain representative configurations. We computed atomic forces from selected snapshots of these dynamic simulations using MLIP-W, as well as empirical (EAM and Tersoff) and ML-based (TabGAP and SNAP) potentials. We then compared these forces with DFT reference forces (see Fig.~\ref{parity_plot_new}). This approach allows us to rigorously test the predictive accuracy of the potentials under realistic simulation conditions and ensure reliable modeling of defect kinetics at finite temperatures.

Fig. ~\ref{parity_plot_new} confirms that MLIP-W provides the most accurate forces at all the three investigated temperatures. Its RMSE increases only modestly from 0.114 eV \AA$^{-1}$ at 300 K to 0.202 eV \AA$^{-1}$ at 1000 K. However, it remains lower than that of any other model at every temperature. The next-best performer, TabGAP, exhibits RMSE values between 0.332 and 0.411 eV \AA$^{-1}$, which is roughly a factor of two higher than MLIP-W in the worst case scenario. SNAP and Tersoff give comparable, larger errors ($\approx 0.40–0.64$ eV \AA$^{-1}$), with Tersoff consistently the less accurate of the two. EAM performs the worst: its RMSE increases from 0.863 eV \AA$^{-1}$ at 300 K to 1.137 eV \AA$^{-1}$ at 1000 K, more than five times the MLIP-W error at that temperature. These trends demonstrate that MLIP-W achieves nearly DFT-level force accuracy, particularly in capturing GB-defect interactions, even at the high temperatures explored in this study.


\subsection{Dynamics and mechanism of defect migration to the GB}
\label{subsec:migration_SIAs}

After comparing structural properties and defect energetics, we now turn to study defect evolution in a W$\langle110\rangle$/W$\langle112\rangle$ GB supercell containing 54,000 atoms. This 
size was chosen as a compromise between computational cost and the ability to reproduce bulk-like conditions within the grains. For the large-scale MD simulations that follow, we restrict the comparison to the two widely used empirical potentials, EAM and Tersoff, which are far less computationally demanding and, to date, have been used more broadly than the more recent MLIPs.

\begin{figure}[t!]
\includegraphics[width=0.95\textwidth]{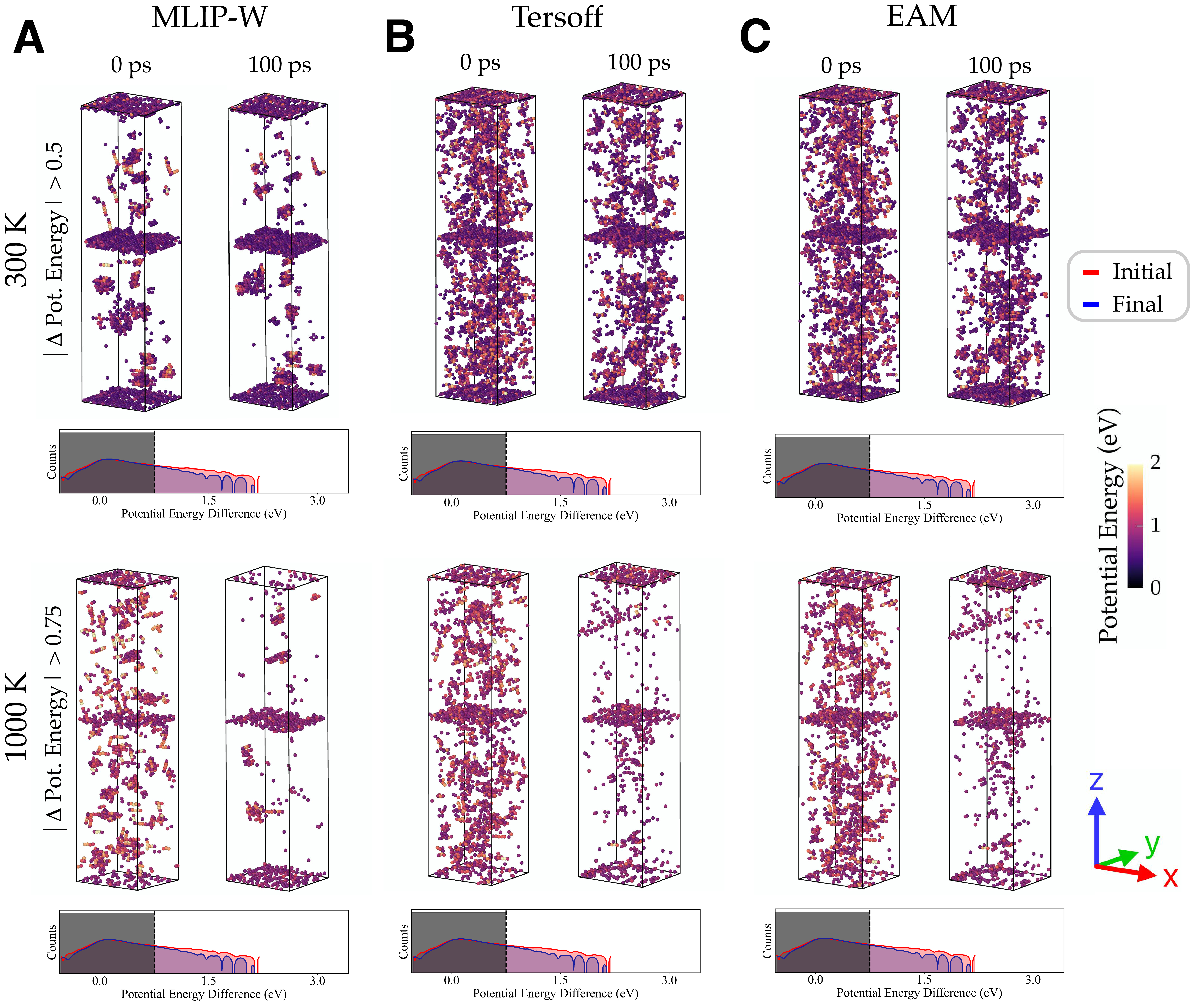}
\centering
\caption{\textbf{Comparison of per atom potential energy differences ($\Delta E_{\text{pot}}$) predicted by the three potentials.} $\Delta E_{\text{pot}}$ distributions and defect configurations for the MLIP-W (A), the Tersoff (B), and the EAM (C) potentials at 300 K (top) and 1000 K (bottom). Each row shows the system at 0 ps and 100 ps after introducing 1,000 defects. Atoms are color-coded by $\Delta E_{\text{pot}}$ relative to the bulk. Bulk-like atoms with energies below 0.5 eV (300 K) and 0.75 eV (1000 K) are not shown. Histograms (bottom regions of each panel) compare $\Delta E_{\text{pot}}$ at 0 ps (red) and 100 ps (blue), with gray regions indicating values below the thresholds. The figure highlights differences in defect evolution across the three potentials. The structure visualizations, including those in this figure and all subsequent ones, were generated using the Open Visualization Tool (OVITO) \cite{ovito}.}
\label{pot_energies}
\end{figure}

To study this process, we introduced 500, 750, and 1,000 defects (Frenkel pairs) into the system to simulate typically expected post-cascade defect densities (hereafter referred to as the cases of low, medium, and high density of introduced defects, respectively), followed by thermalization at the appropriate temperatures (300\,K, 650\,K, and 1000\,K) for 8\,ps, a value that falls within the range commonly employed for this type of simulation.
Fig.~\ref{pot_energies} shows the initial (0\,ps) and final (100\,ps) configurations for the third case, i.e., high density of defects, obtained using MLIP-W (Fig. \ref{pot_energies} A), the Tersoff potential (Fig. \ref{pot_energies} B), and the EAM potential (Fig. \ref{pot_energies} C), at 300\,K (top panels) and at 1000\,K (bottom panels). Atoms in Fig.~\ref{pot_energies} are color-coded according to their potential energy difference (\(\Delta E_{\text{pot}}\)) relative to the bulk for each potential. We only show atoms with energies above 0.5\,eV at 300\,K and 0.75\,eV at 1000\,K, in order to highlight defected or strained regions. At high temperatures, the increase in the minimum energy value accounts for thermal vibrations. In addition to the atomic configurations, histograms in Fig.~\ref{pot_energies} illustrate the distribution of potential energies. In these histograms, the red curve corresponds to the initial distribution (0\,ps), while the blue curve represents the final distribution (100\,ps). Moreover, atoms not visualized in the atomic configurations correspond to those within the gray shaded area of the histogram.

At 300 K, the MLIP-W (Fig. \ref{pot_energies} A) effectively organizes atoms with a \(\Delta E_{\text{pot}}\) larger than 0.5 eV into well-defined regions already at 0 ps, corresponding to defect-rich areas 
near the GB. This indicates that the potential 
stabilizes defects into low-energy, structured configurations, 
such as dumbbells, characteristic of self-interstitials in BCC 
metals. Few atoms exhibit \(\Delta E_{\text{pot}} > 2\) 
eV, suggesting that the potential avoids energetically 
excessive configurations and efficiently relaxes local 
stresses. At 100 ps, the number of high-energy atoms (\(\Delta 
E_{\text{pot}} > 0.5\) eV) slightly decreases from 3.7\% to 
3.4\%, due to stress relaxation and the migration of high-energy atoms to the GB, where defect formation energies are 
lower. This migration is facilitated by the initial 
organization of SIAs into low-energy \(\langle111\rangle\) dumbbell configurations. This configuration enables the defect to migrate by sequentially displacing neighboring atoms (i.e., via the crowdion mechanism \cite{was2016fundamentals}), thereby moving toward the GB, where its presence is energetically more favorable since the atomic packing density is lower. In contrast, the Tersoff potential (Fig. \ref{pot_energies} B) leads to a disordered distribution of atoms where 8.0\% of them have  \(\Delta E_{\text{pot}} > 0.5\,\mathrm{eV}\) at 0 ps, more than twice those of the MLIP-W. These high-energy 
atoms are randomly scattered. No structured defect 
configurations like dumbbells are observed, indicating that the Tersoff 
potential is limited in capturing defect energetics and 
stability. This atomic disorder suggests hindered migration of 
interstitial defects to the GB, hampering defect annihilation. After 100 ps, the number of high-energy atoms decreases to 6.1\%, but the disordered distribution 
persists. Finally, the EAM potential (Fig. \ref{pot_energies} C) leads to 7.0\% of atoms exhibiting \(\Delta E_{\text{pot}} > 0.5\,\mathrm{eV}\) at 0 ps. This number is similar to that estimated with the Tersoff potential. However, unlike the Tersoff potential, some order is seen, with signs of crowdion-like structures forming. A significant portion of atoms exhibits \(\Delta E_{\text{pot}} > 2\,\mathrm{eV}\), indicating the inability of the potential to relax local stresses during thermalization. After 100 ps, the number of atoms with high-energy values decreases to 5.4\%, but a significant fraction remains, meaning that the EAM potential is limited in modeling defect relaxation and migration. Moreover, the broadness of the tail in the energy distribution is much larger than that for the MLIP-W and Tersoff potentials.

At 1000 K, the MLIP-W (Fig. \ref{pot_energies} A) maintains a high level of structural order at 0 ps, with atoms having \(\Delta E_{\text{pot}}>0.75\) eV located in definite defect-rich regions, mainly forming dumbbell configurations. The percentage of these atoms is 3.0\%, slightly lower than at 300 K. The number of atoms with \(\Delta E_{\text{pot}} > 2\)  eV increases, probably due to the enhancement of thermal vibrations. As shown in the lower part of Fig. \ref{pot_energies}, after 100 ps, the migration of SIAs to the GB becomes more efficient and the number of atoms mainly located at the GB with \(\Delta E_{\text{pot}} > 0.75\) eV decreases to 1.4\%. This defect annihilation, reflected in the histogram, demonstrates the ability of the potential to capture defect dynamics at elevated temperatures and is consistent with experimental observations of denuded zones near GBs \cite{ELATWANI2018206}. Similarly to the situation at 300 K, calculations with the Tersoff potential (Fig. \ref{pot_energies} B) show that the system remains disordered at 1000 K with no clear defect structures such as dumbbells. The remaining atoms do not correspond to defected regions, indicating that the Tersoff potential leads to atomic configurations in which defect-related and thermal displacement environments overlap significantly, which complicates defect identification. In the final state, only 1.8\% of the atoms exhibit \(\Delta E_{\text{pot}} > 0.75\) eV, most of which are located at the GB. However, this result should be viewed with caution since, as mentioned just above, high potential energy does not reliably indicate an actual defect. Finally, calculations with the  EAM potential  (Fig. \ref{pot_energies} C) indicate that high-energy atoms remain localized in defect regions, with 2.6\% of atoms exhibiting \(\Delta E_{\text{pot}} > 0.75\) eV. This number of atoms is nearly double that calculated with the MLIP-W. The EAM potential also retains some crowdion-like structures very close to the GB, which is unexpected given their proximity. The superior performance of MLIP-W likely originates from its extensive DFT-based training dataset. This enables MLIP-W to accurately describe stable defect configurations, such as $\langle111\rangle$ interstitial dumbbells, as well as subtle transition pathways and energy barriers between intermediate states. By contrast, empirical potentials typically lack explicit fitting to these non-obvious energetic features, resulting in a less accurate, rougher PES. Such discrepancies can lead to misinterpretations in radiation-induced damage studies, especially when precise modeling of defect sinks is necessary to predict material performance.

\begin{figure}[t!]
\includegraphics[width=0.95\textwidth]{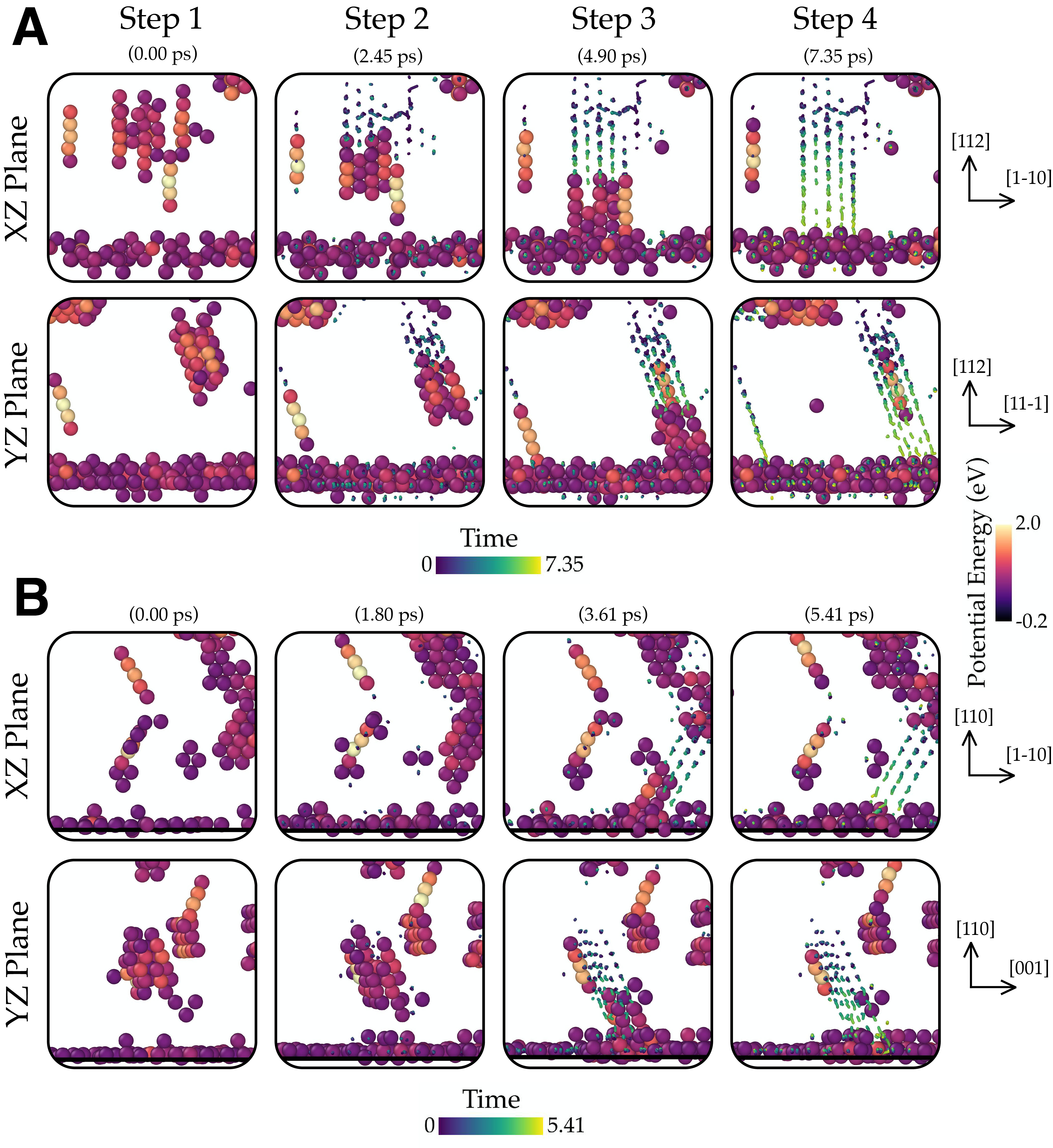}
\caption{\textbf{Migration pathways of SIAs toward GBs.} Time evolution snapshots of $\langle111\rangle$ dumbbell migration from (A) $\langle112\rangle$ and (B) $\langle110\rangle$ grain interiors to the GB. 
Each column represents a time step, with [112] (top) and [110] (bottom) grain 
views. Only atoms with \(\Delta E_{\text{pot}} > 0.5\) eV from bulk are shown, 
colored by their energy values. Migrating atoms leave color-coded trails (bottom bar) indicating their positions over time. Crystallographic directions are shown on the right.}
\label{migration}
\end{figure}

The results reported above are supported by direct visualizations from the MD movies provided in Supplementary Note 2, which show the dynamics captured in 40 snapshots over 100 ps for each IP (see Figs. S2-S10).
The time-evolution snapshots shown in Fig.~\ref{migration} A and B illustrate the dynamic migration of $\langle111\rangle$ dumbbells from both the $\langle112\rangle$ (Fig.~\ref{migration} A) and  $\langle110\rangle$  (Fig.~\ref{migration} B) grain interiors to the GB at 300 K, respectively. These snapshots show that the stable dumbbell configurations facilitate an efficient defect transport mechanism in which SIAs undergo sequential displacements along preferential crystallographic directions, effectively propagating defects toward the GB, as previously mentioned. The top and bottom rows, in the same figure, show cross sections along \{11$\overline{1}$\} (XZ plane) and \{1$\Bar{1}$0\} (YZ plane), providing complementary perspectives on this migration mechanism. The crystallographic directions shown on the right side of Fig.~\ref{migration} allow direct correlation with the underlying lattice. This efficient transport process, in which defects are directed to the GB as an effective sink, highlights the ability of the MLIP-W to accurately capture the mechanism of defect evolution, in agreement with DFT and Monte Carlo predictions \cite{VALLES2017277}. 

In addition, for a more objective analysis, we define a region around the GB whose \textit{z} boundaries do not intersect any atomic layers, thus preventing thermal fluctuations from being mistaken for actual migrations. By comparing the atoms initially in this region with those at the end of the simulation, we identify new atoms (tracked via their unique IDs) as those that have migrated to the GB. The exact values are shown in Table S1 of Supplementary Note 3. At low temperatures (e.g., 300 K), differences among the three IPs become particularly pronounced: the number of SIAs that migrate to the GB, as calculated using the MLIP-W, is approximately ten times higher than the number calculated using the Tersoff and EAM IPs. The results provided by the MLIP-W agree much better with previously reported DFT data, which have shown that the migration barriers for SIAs in tungsten are close to zero \cite{BECQUART200723}, and with Monte Carlo simulations predicting that this migration can occur at temperatures as low as 5-10 K \cite{VALLES2017277}. They are also in agreement with experimental data reported for nanostructured W samples sequentially implanted with C, which show that these materials exhibit greater H retention than their bulk counterpart, as the SIAs migrate to the GB, leaving vacancies in the grain interior where H is accumulated \cite{GonzalezArrabal2020,Panizo2019}. 
At intermediate and high temperatures, the differences between the potentials become subtler. At 650\,K, the number of defects that have migrated to the GB calculated with MLIP-W and Tersoff potentials is very similar on average, while that calculated with the EAM potential is a factor of two lower. At 1000 K, the number of defects that have migrated to the GB calculated with the MLIP-W is about 2.5 times higher than that calculated with the Tersoff or the EAM potentials. These results prove the high fidelity of the MLIP-W in capturing both the energetics and kinetics of SIAs migrating to GB, as well as the limitations of the Tersoff and EAM IPs in describing these processes.

\subsection{GB structure after defect migration}
\label{subsec:structure_GB}

\begin{figure}[t!]
\includegraphics[width=0.85\textwidth]{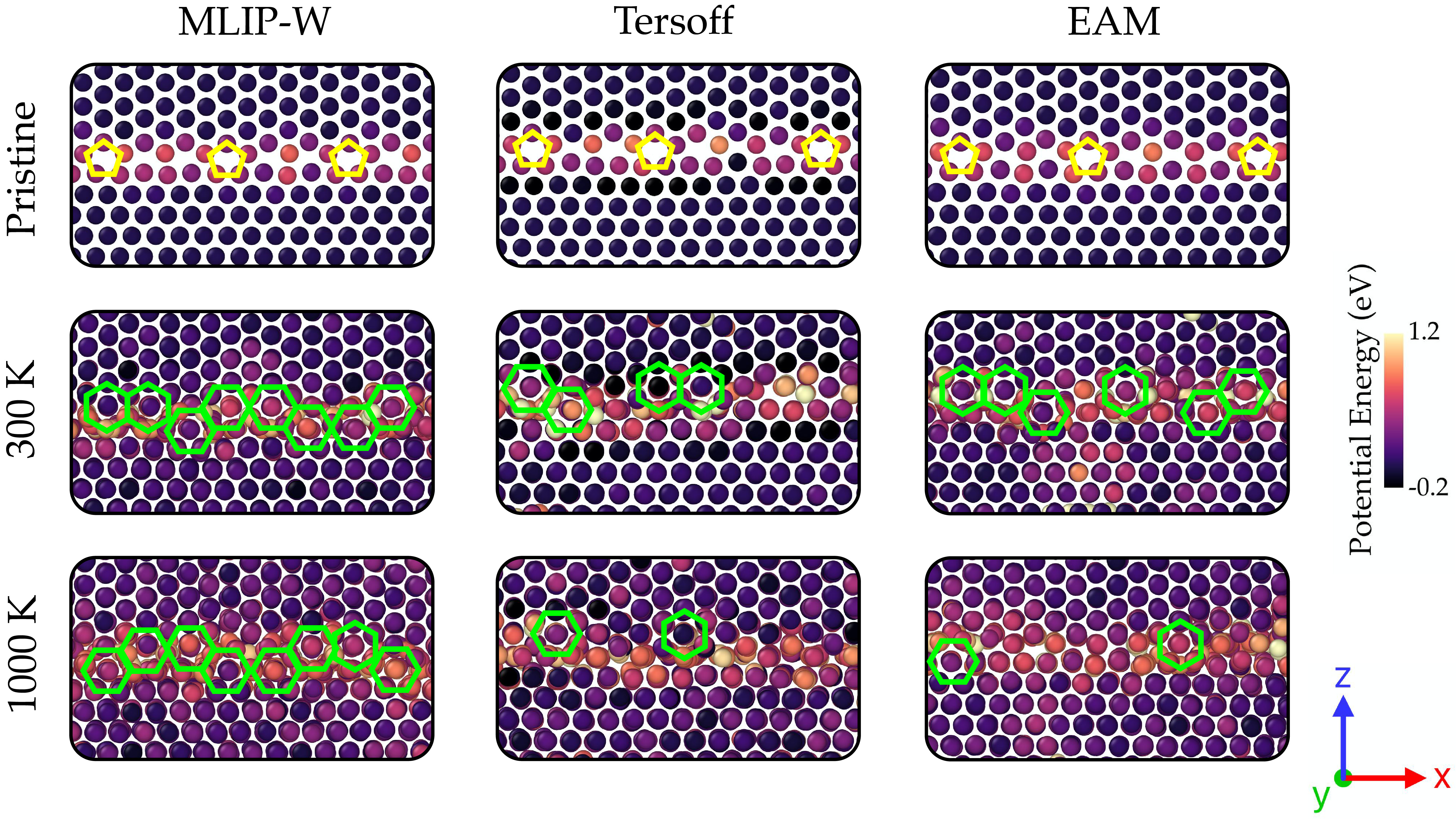}
\centering
\caption{\textbf{Atomic structures and stability of GBs.} GB structures with the MLIP-W (left), the Tersoff (center), and the EAM (right) potentials for pristine (top), 300 K (middle), and 1000 K (bottom) conditions. Atoms are color-coded by their potential energy difference relative to the bulk (see legend). In the 0 K case, recurring GB motifs are marked by yellow pentagons, while at 300 K and 1000 K, distinct GB structures are highlighted by green hexagons. These patterns reveal the thermal and IP-dependent stability of the GB (see text).}
\label{GB_potE}
\end{figure}

Having determined the number of defects migrating to the GB, we now deepen into the atomistic structure of the interface as calculated each IP. Fig.~\ref{GB_potE} shows the potential energies of the GB atoms relative to the bulk as determined with each potential. At 0 K, for pristine samples (top row), all three potentials converge to a similar interface structure with a periodic pentagonal motif (yellow) and a comparable equilibrium geometry. However, the maximum GB-bulk energy difference depends on the selected IPs being 0.77 eV (for the MLIP-W), 0.91 eV (for the EAM potential), and 1.05 eV (for the Tersoff potential). Based on previous DFT studies \cite{Suarez-Recio24,Gonzlez2019}, the interface naturally forms GB-aligned grooves, which serve as defect sinks. At temperatures from 300 K to 1000 K, the GB structure estimated with the MLIP-W remains almost unchanged, while the motifs begin to blur when the other IPs are used (not shown in Fig.~\ref{GB_potE}). When we further analyze the GB after introducing Frenkel pairs (high density case) and allow the system to evolve at different temperatures, diverse changes in defect behavior are observed. The middle (300\,K) and bottom (1000\,K) rows of Fig.~\ref{GB_potE} show the final GB configurations under defect implantation conditions comparable to those in Fig.~\ref{pot_energies}. The DFT findings mentioned above indicate that SIAs arrange in rows along these grooves because these regions have a lower defect formation energy than the bulk, mainly due to the larger space available for accommodation. For the MLIP-W, SIAs are observed to effectively reach the GB and align along these grooves. In particular, at finite temperatures, atoms near both the W$\langle112\rangle$ and W$\langle110\rangle$ layers adjacent to the SIAs row shift into slightly arc-shaped crowdions to minimize the energy of the system. These regions are marked as green hexagons in Fig.~\ref{GB_potE} to illustrate the characteristic atomic packing. This phenomenon is explained in more detail in Supplementary Note 4, where Fig. S11 clearly illustrates the formation of the crowdion. At low temperatures, such distribution is also captured, up to a certain degree, by the Tersoff and the EAM IPs. However, the ability of these empirical IPs to reproduce the GB structure decreases with increasing temperature to such an extent that they predict a compromised integrity of the GB, which is unexpected at the temperatures studied (up to 1000 K).

This contrast in high-temperature robustness underscores the significance of training data. During the active learning approach, the MLIP-W accurately learned from \textit{ab initio} data in these complex environments, capturing how defects are accommodated at the GB, thanks to a complete and rigorous description of the PES.

\subsection{Descriptor-based analysis of defects}
\label{subsec:analysis_and_quant}

To further validate and analyze the performance of the MLIP-W in describing defect evolution at W$\langle110\rangle$/W$\langle112\rangle$ GBs, we used the Fingerprinting and Visualization Analyzer of Defects (FaVAD) \cite{VONTOUSSAINT2021107816}. This approach is based on DVs and allows us to characterize defects based on their local atomic environments.
Although the DeePMD-kit \cite{WANG2018178,10.1063/5.0155600,LU2021107624} generates its own DVs, their high dimensionality made them impractical for our 54,000-atom cell, and we instead use the multi-body descriptor framework called ‘Smooth Overlap of Atomic Positions’ (SOAP) \cite{Bartok2013} to compute a reduced set of descriptors (see Section~\ref{subsec:FaVAD}).
In this fingerprinting method, we measure the Euclidean distance, denoted as \( d^M(T) \), between the DV of each atom in the defect sample, \( \vec{q}^{\,i} \), and the mean DV of a defect-free (bulk) reference configuration at a certain temperature \(T\), \( \vec{v}(T) \). This distance difference between two local atomic environments forms the basis for a probabilistic interpretation of the results, expressed as \( p\bigl(\vec{q}^{\,i} \mid \vec{v}(T)\bigr) \). We find that the distribution of these normalized distances is well approximated by a chi-squared distribution, \(\chi^2_K\), with \(K\) degrees of freedom. Mathematically, this is expressed as

\begin{equation}
p\bigl(\vec{q}^{\,i} \mid \vec{v}(T)\bigr) 
= \frac{\exp\Bigl[-\tfrac{1}{2}\,d^M(T)^2\Bigr]\,\Bigl(d^M(T)\Bigr)^{\tfrac{K}{2}-1}}
{2^{\tfrac{K}{2}}\,\Gamma\!\Bigl(\tfrac{K}{2}\Bigr)}
\,,
\end{equation}
where \(\Gamma\) is the Gamma function. In our analysis, as observed in Fig. S13, \(K=4\) provides an optimal fit to the distributions \cite{VONTOUSSAINT2021107816}. This probabilistic framework allows defining a temperature-dependent threshold to differentiate genuine defects from thermal oscillations. To establish this threshold, first, the bulk reference DVs are computed at each simulation temperature. Since defects in BCC materials predominantly form \(\langle111\rangle\) dumbbell configurations \cite{10.1063/5.0048740}, reference DVs for thermally equilibrated interstitials are also determined. Note that reference DVs are determined individually for each potential. By mapping distances relative to the bulk and interstitial reference DVs for a pure bulk configuration at a given temperature, a distribution that defines the bulk region in 2D space is obtained. Thus, when analyzing configurations containing defects or GBs, points outside this region are considered defects. This 2D methodology represents an advance over that used in Ref. \cite{VONTOUSSAINT2021107816}, as it allows for a more accurate identification of crowdion-like interstitials, which are the most likely structures. Further details of this methodology, including how the area thresholds are determined, are given in Supplementary Note 5. In particular, Fig. S12 shows the distributions found for a 54,000-atom bulk system thermalized at each temperature and for each potential. The DV distributions reveal that while the MLIP-W broadens uniformly at higher temperatures, empirical IPs—particularly the Tersoff IP—develop bifurcated structures at 1000 K that are indicative of distinct structural motifs, a behavior likely attributable to an artifact of the potential. Exact thresholds can be found in Table S2 of the Supplementary Note 6.

\begin{figure}[t!]
\includegraphics[width=0.95\textwidth]{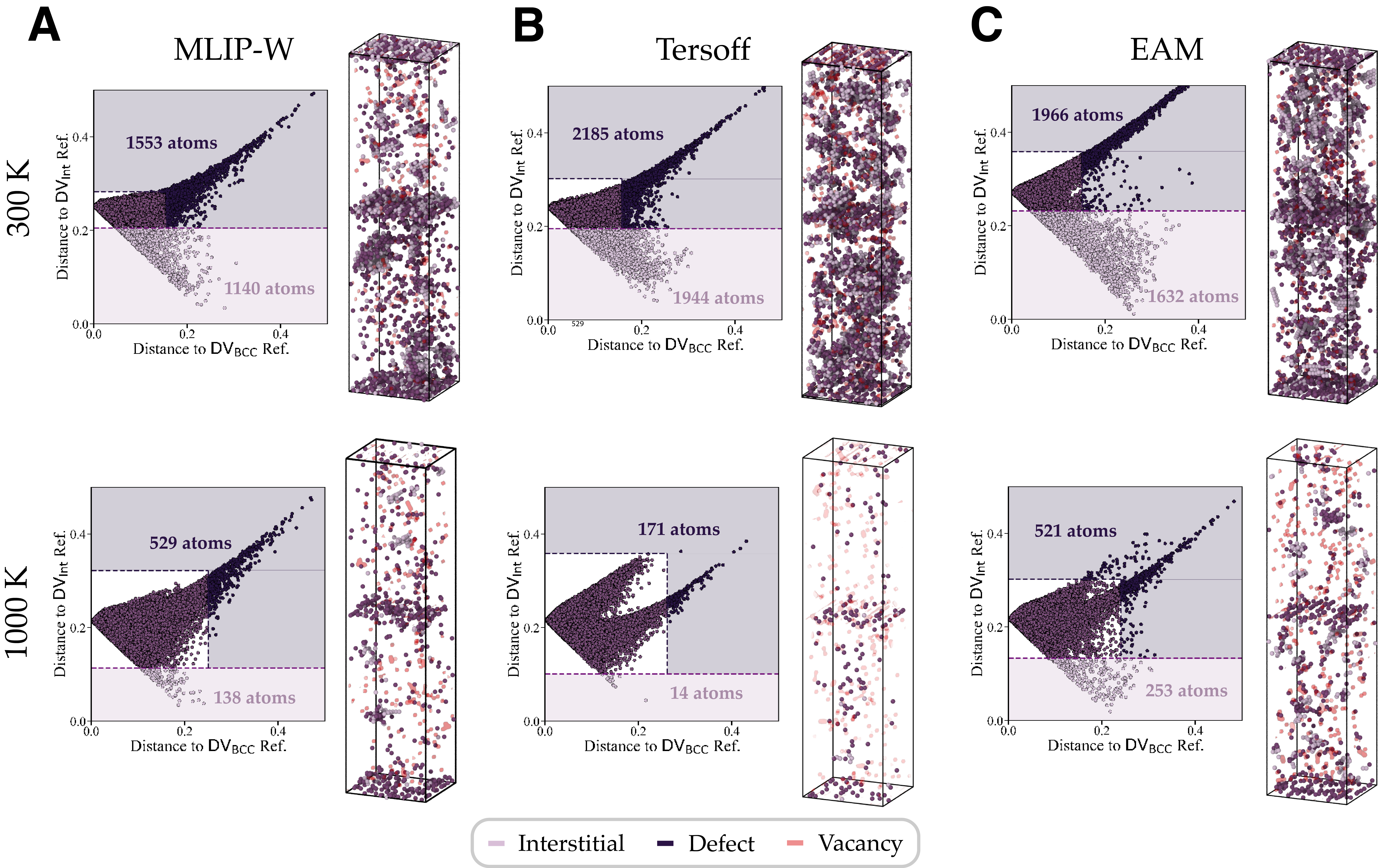}
\centering
\caption{\textbf{Descriptor-based defect classification in W GB.} Distribution of DV distances and corresponding 54,000-atom W supercell for the high density of introduced defects at 300 K (top) and 1000 K (bottom) with the MLIP-W (A), Tersoff (B), and EAM (C) potentials. Left panels show distances to bulk (horizontal) and interstitial (vertical) reference DVs, highlighting distinct “sheets” for interstitials (lower) and close to the GB defects (upper). Right panels depict atoms color-coded by environment: interstitials (light purple), vacancies (orange), and unidentified defects (dark purple). Bulk atoms are omitted for clarity.}
\label{analysis-total_new2}
\end{figure}

Once the bulk region is defined, atoms with DV values below the lower threshold are identified as dumbbell-like interstitials (see Fig. S12 for clarification). The remaining atoms—those near vacancy pairs, non-dumbbell-aligned interstitials, and atoms belonging to the GB—are grouped as unidentified defects. While a more detailed classification based on specific reference DV values for each defect type is feasible, such analysis is not needed for the goals of this study. On the other hand, identification of single vacancies and void formation was performed using Voronoi-based tessellation methods, as implemented in OVITO and described in previous works \cite{Wyszkowska2025Nanoscale,Ustrzycka2024Atomistic}. This simple and robust method enables a quantitative estimation of local atomic volumes by dividing space into Voronoi polyhedra, thereby allowing detection of undercoordinated atoms and volumetric anomalies indicative of vacancy clusters or voids. Fig.~\ref{analysis-total_new2} shows DV distributions (left) and simulation cells (right) for the high density of introduced defects case, both at 300 K and 1000 K. Atoms are color coded by environment: SIAs (light purple), vacancies (orange), 
and unidentified defects (dark purple), with bulk atoms 
omitted for the sake of clarity. Two new "sheets" emerge: the lower one 
for SIAs and the upper one for the GB and unidentified 
defects. In Fig. S13 of Supplementary Note 7, a detailed comparison of the DV distributions for bulk, pristine GB, and defected GB, each thermalized at 300 K, is presented. At 300 K, all 
potentials exhibit similar distributions, though the EAM IP shows a 
slight “V” shape. The MLIP-W produces fewer SIAs and 
defects, resulting in fewer points in the 3D cell. The EAM potential behaves similarly, whereas configurations produced with the Tersoff potential, as previously stated in Subsection~\ref{subsec:migration_SIAs}, show substantial overlap between defect and thermally induced environments, making a clear defect identification difficult.
At 1000 K, differences become more pronounced. The distribution found with the MLIP-W expands like in the bulk case, sharply reducing 
interstitials (by a factor of ten) and unidentified defects 
(by a factor of three) due to a broader bulk region filtering 
thermally vibrating atoms. The Tersoff IP shows an even steeper decline, from 4,129 defects at 300 K to only 185, although these numbers should be interpreted cautiously. EAM behaves 
similarly to the MLIP-W but introduces a third “sheet,” suggesting 
additional defective regions, i.e., atoms whose local environments deviate from the bulk and differ from the previous defect environments observed at lower temperatures. It also retains twice as many 
SIAs as the MLIP-W (253 vs. 138), likely due to fewer 
migration events.

\subsection{Quantifying the temperature dependence of defect migration}
\label{subsec:temp_dep}

\begin{figure}[t!]
\includegraphics[width=\textwidth]{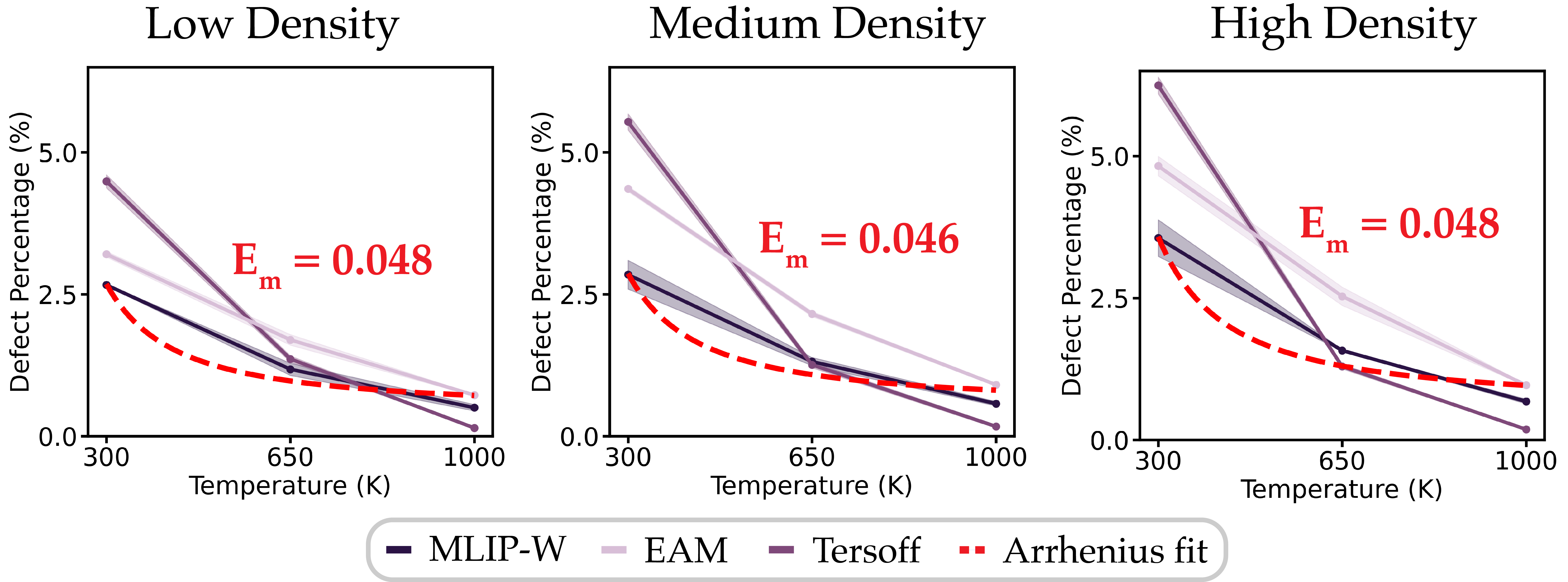}
\centering
\caption{\textbf{Temperature dependence of defect counts in the grain interior.} Percentage of defects in the grain interior (excluding the GB region) as a function of temperature for each potential. Results are shown separately for three initial defect densities, namely, low, medium and high, respectively. In each panel, the dashed red line represents a fit to an Arrhenius form for the MLIP-W.}
\label{curves_defects}
\end{figure}

Fig.~\ref{curves_defects} shows the average number of defects calculated with each IP as a function of temperature and initial defect density. Here, defects annihilated either by recombination with vacancies or migration into the GB region (treated as a defect sink) are not included in the final count. Full numerical values are provided in Supplementary Note 8, where Tables S3, S4, and S5 list the defect counts for the MLIP-W, the Tersoff, and the EAM potentials, respectively. At 300 K, in agreement with our previous observations, the MLIP-W consistently yields the lowest number of defects, whereas the Tersoff and the EAM potentials predict higher counts. Specifically, for the low density of introduced defects case, the number of defects calculated with the Tersoff and the  EAM potentials is 1.7 and 1.2 times larger than that calculated with the MLIP-W, respectively. For the medium density of introduced defects case, the number of defects calculated with the Tersoff and the EAM IPs is 1.7 and 2.3 times higher than that calculated with the MLIP-W. Finally, for the high density of defects introduced case, the number of defects calculated with the Tersoff and the EAM IPs is 1.8 and 1.4 times larger than that calculated with the MLIP-W. These trends indicate that, at low temperatures, both the Tersoff and the EAM IPs yield more residual defects in the grain interior than the MLIP-W, evidencing the superior ability of the MLIP-W to capture defect migration and subsequent annihilation and/or reorganization at the GB. As temperature increases, the differences between the IPs become less pronounced. The EAM potential appears to follow a trend similar to the MLIP-W, albeit with consistently higher defect counts, whereas the slope of the Tersoff curves is considerably steeper, likely reflecting its inability to accurately capture defect behavior at high temperatures, resulting in defect counts even lower than those predicted by the MLIP-W. 

A natural way to validate the temperature dependence of these defect count curves is to interpret them in the light of the kinetic rate theory by Arrhenius \cite{Arrhenius1889,was2016fundamentals}. However, it is important to note that, as previously reported, deviations from Arrhenius behavior may occur at significantly higher temperatures than those considered in this study (i.e., temperatures ranging from approximately 60 to 75\% of the melting temperature). At these high temperatures, the migration regime of interstitial defects transitions from discrete, infrequent hops to continuous Brownian-like motion driven by interactions with thermal phonons. This phenomenon takes place because the activation energy for SIA migration is so low (comparable to or smaller than thermal energy) that the defects do not remain localized for extended periods of time. Instead, they undergo continuous migration, resulting in a non-Arrhenius, linear temperature dependence of their diffusion coefficient at elevated temperatures \cite{Dudarev2008Arrhenius,Zhang2025Ab}.

Nonetheless, in our study, the highest temperature explored (1000 K) is only about 27\% of the melting temperature of tungsten (3695 K), falling well within the expected Arrhenius regime. For this reason and assuming that SIAs, once formed, diffuse to the GBs, that act as sinks for them, the annihilation rate or net decrease of SIAs at the GB with regard to the grain interior can be considered an effective measure of their migration. Such annihilation rate $v$ follows a classical Arrhenius form,

\begin{equation}   
v = \frac{\Delta n_m}{N \Delta t} \propto \exp\!\Bigl(\frac{S_m}{k_B}\Bigr) \exp\!\Bigl(-\frac{E_m}{k_B T}\Bigr),
\label{equation_arr}
\end{equation}

\parindent 0 em

where $\Delta n_m$ is the variation in the number of defects inside the grain (excluding vacancies) in a time $\Delta t$, \(N\) is the total number of atoms, \(S_m\) is the migration entropy, \(E_m\) is the migration energy, and \(k_B\) is the Boltzmann constant. The temperature-independent prefactor \(\exp\bigl(S_m/k_B\bigr)\) represents the entropic contribution to the formation of an activated state, while \(\exp\bigl(-E_m/k_B T\bigr)\) captures the energetic barrier to defect migration. Considering that $\Delta t$ is constant and equal to 100 ps in all simulations, we can fit Eq.~\eqref{equation_arr} to our $\Delta n_m/N$ vs. $T$ data to extract \(E_m\).
We find that the MLIP-W yields a migration energy of \(0.048 \pm 0.001\) eV, which is in good agreement with the typical \(\sim 0. 05\) eV range reported for SIAs in W—for instance, 0.054 eV \cite{was2016fundamentals} or 0.046 eV \cite{Amino01022011}. For the EAM potential, the fits lead to a slightly lower migration energy value of \(0.043 \pm 0.001\) eV. The migration energy as fitted with the Tersoff potential has a much larger value than that calculated by the other potentials (\(0.082 \pm 0.008\) eV). These results are additional evidence of the inaccuracy of the Tersoff potential to  distinguish defects from thermal noise. 
The good agreement of \(E_m\) calculated with the MLIP-W and the reference data highlights its ability to capture defect kinetics and GB-mediated self-healing mechanisms in W under conditions where previously reported empirical IPs do not perform well.

\parindent 1 em

\section{Discussion}

In this work, we have developed an IP, dubbed MLIP-W, which accurately describes defect migration in the vicinity of the W$\langle$110$\rangle$/W$\langle$112$\rangle$ GB, commonly observed in experiments.
By integrating an active learning approach with a diverse training set—spanning bulk, GB, defect-rich, and thermally perturbed configurations—the MLIP-W provides quantum-level energy accuracy, capturing the key features of the DFT-derived PES while supporting large-scale MD simulations. Our model not only outperforms the widely used EAM and Tersoff empirical IPs, but also more recently developed ML potentials, TabGAP and SNAP, in reproducing critical properties such as lattice parameters, vacancy formation energy, and second-neighbor di-vacancy binding energy, matching DFT and experimental benchmarks (see Table \ref{comparison_pot}). Crucially, the MLIP-W captures the negative binding energy for second-neighbor vacancies, a feature essential for modeling aggregation but absent in traditional IPs. Moreover, we have validated the MLIP-W model using the same type of defect-rich GB configurations and target finite temperatures as in our production runs. As Fig.~\ref{parity_plot_new} shows, MLIP-W is the most accurate potential by far, with force RMSEs remaining below 0.21 eV Å$^{-1}$ up to 1000 K. All other models, on the other hand, exhibit substantially larger deviations, particularly EAM. This benchmark confirms the reliability of MLIP-W for the reported simulations.

Once validated, we used MLIP-W, together with the empirical IPs (cheaper and more widely used than the other ML potentials), in large-scale MD simulations. We found that MLIP-W rapidly arranges self-interstitials into stable $\langle111\rangle$ dumbbells, allowing efficient migration to GBs via a crowdion mechanism. This is in contrast with the disordered defect distributions predicted by empirical potentials. Another important point to consider is the structural integrity of the GB. In this regard, the MLIP-W is the only IP among those studied that preserves the structural integrity of the GB over the temperature range studied in this work. Our observations using the MLIP-W reveal that when SIAs migrate from the grain interior to the GB, they preferentially align along the GB grooves, which is consistent with previous DFT studies on this particular GB \cite{Suarez-Recio24,Gonzlez2019}. Once positioned, these SIAs form an arc-shaped crowdion configuration with atoms from the outermost layer of the grain (observed for both W$\langle$110$\rangle$ and W$\langle$112$\rangle$ orientations), displacing a GB layer atom into an adjacent groove. While the empirical IPs can partially reproduce these mechanisms at low and medium temperatures, they struggle to reproduce stable configurations at elevated temperatures (up to 1000 K), leading to unphysical grain decohesion that is inconsistent with experimental expectations in this thermal regime. The superior performance of our model stems from its training strategy, in which MLIP-W learns the local PES with DFT accuracy. The active-learning protocol systematically targets the high-uncertainty regions of the PES. This allows the model to capture the energy minima of diverse, complex environments, as well as the subtle transition pathways and energy barriers that connect intermediate states.

Moreover, our descriptor-based analysis reveals that configurations generated with MLIP-W distinctly differentiate between defect-related environments and thermal fluctuations. In contrast, configurations obtained with traditional IPs, particularly the Tersoff model, exhibit substantial overlap between defect-induced and thermally driven atomic displacements. This is consistent with the blurred separation observed in the per-atom potential energy analysis. Finally, an Arrhenius analysis of defect annihilation rates yields a migration energy of 0.048 eV when using the MLIP-W, in agreement with the reported value range \cite{was2016fundamentals,Amino01022011}. 

The results obtained in this work underscore the usefulness of MLIP as a tool for simulating self-healing and defect-GB interactions in  W GBs. Future work will extend this framework to investigate the role of light impurity atoms (He, H) in self-healing processes in nanostructured W, paving the way for understanding material degradation in realistic and highly complex nuclear fusion environments.

\section{Methods}
\label{sec:Methods}

\subsection{MLIP-W development}
\label{subsec:Training_DeePMD}

The MLIP-W potential was developed using the Deep Potential-Smooth Edition (DP-SE) framework within the DeePMD-kit~\cite{WANG2018178}. The entire fitting procedure consists of four main steps, namely training, exploration, labeling, and production. First, initial training data to feed the deep neural network were generated from two fully relaxed DFT configurations: a pristine 5×5×5 BCC tungsten supercell (250 atoms) and a W⟨110⟩/W⟨112⟩ GB system (456 atoms). More details on the construction of this GB can be found in section~\ref{subsec:DFT_calculations}. 

To sample the PES, 100 perturbed configurations for each system were generated by using random atomic displacements (\(\delta R_{\text{max}} = \pm0.1\)). In addition, defect configurations were incorporated by introducing isolated vacancies and SIAs at random positions, ensuring a minimum spacing of 3 Å to avoid lattice overlap. This approach was adopted to simulate irradiation environments by capturing the diverse local atomic configurations that may occur under such conditions. Note that four potentials were trained independently with the same hyperparameters but different random seeds, yielding an ensemble that will be used in the next step. These models used smooth DVs truncated at a cutoff radius of 6\,Å, with a smoothing function applied beyond 3\,Å. The descriptor network consisted of three layers of 25, 50, and 100 neurons, while the fitting network used three layers of 120 neurons each. Training was performed with an initial learning rate of 0.002, which decayed exponentially after 5,000 steps. A total of 2,000,000 steps were performed for convergence. The energy, force, and virial components were weighted with dynamic prefactors during training. The energy prefactor ranged from 0.02 to 1 and the force prefactor decreased from 1,000 to 1. 

In the exploration step, an iterative active learning protocol was implemented to refine the potential. Using the models described above, short MD simulations (on the order of a few tens of picoseconds) in the NVE ensemble were performed at temperatures of 300 K, 650 K, and 1000 K, corresponding to the three different regimes identified by Beyerlein et al.~\cite{BEYERLEIN2013443}, each with a time step of 5 femtoseconds and 10,000 MD steps. In each cycle, 10 Frenkel pairs were introduced  with a separation of \(\geq 3\) Å, allowing for natural defect evolution near GBs. The aforementioned potential ensemble was then used to measure the maximum standard deviation of the predicted atomic forces of the samples obtained by MD, i.e.,

\[
\epsilon_t = \max_i \sqrt{ \left\| F_{w,i}(R_t) - \langle F_{w,i}(R_t) \rangle \right\|^2 },
\]

where \(\epsilon_t\) and \(R_t\) denote the maximum force deviation and the atomic coordinates of a given structure \(t\), respectively. The index \(i\) runs over all atoms in the structure and the index \(w\) denotes the \(w\)th potential model in the ensemble, meaning that the atomic forces \(F_{w,i}\) are ensemble averaged. Four potentials are often considered sufficient to construct this ensemble. The above equation was then used as a criterion to select candidate structures from the MD trajectories, i.e., those structures whose \(\epsilon_t\) satisfied the inequality \(\sigma_f^{\text{min}} < \epsilon_t < \sigma_f^{\text{max}}\) were selected and sent to the next step. In practice, \(\sigma_f^{\text{min}}\) and \(\sigma_f^{\text{max}}\) are set empirically, with no fixed rule, just to ensure that a reasonable fraction of candidates is selected in each iteration. In this work, \(\sigma_f^{\text{min}} = 0.20\) and \(\sigma_f^{\text{max}} = 1.00\) were used.

Third, the chosen candidates were recalculated by DFT in the labeling step, with the parameters described in Sect. \ref{subsec:DFT_calculations}. The computed (i.e., labeled) samples were fed back into the training step, triggering the next iteration of the DeePMD scheme. The first three steps were performed iteratively until DeePMD converged, i.e., no more candidates were found under the \(\epsilon_t\) criterion. Finally, the model was compressed after the potential was generated. In this context, compression refers to the use of techniques such as tabulated inference, operator merging, and precise neighbor indexing to streamline computations and reduce memory overhead, thereby enhancing performance. The results presented in this paper have been tested on this compressed potential and they embody the final production step.

\subsection{DFT calculation details}
\label{subsec:DFT_calculations}

The Vienna \textit{Ab initio} Simulation Package (VASP)~\cite{PhysRevB.47.558,kresse1996efficient,kresse1999ultrasoft} was employed to perform DFT calculations using the projector augmented wave (PAW) method~\cite{PhysRevB.50.17953} in combination with the generalized gradient approximation (GGA) for the exchange-correlation functional, as formulated by Perdew-Burke-Ernzerhof (PBE)~\cite{PhysRevLett.77.3865}. These calculations were conducted to generate the training dataset and to establish property benchmarks. The W pseudopotential considers 6 valence electrons in the 5d$^4$6s$^2$ configuration.

Two representative systems were fully relaxed using DFT: (i) a 5 × 5 × 5 supercell of bulk W, containing 250 atoms, to capture the intrinsic properties of the BCC lattice; and (ii) the W$\langle110\rangle$/W$\langle112\rangle$ interface, with 456 atoms in a cell of dimensions 31.09 Å × 10.98 Å × 31.72 Å. The construction of the W$\langle110\rangle$/W$\langle112\rangle$ interface follows the methodology described in Ref. \cite{González_2015}, where two 6-layer slabs are joined along the $z$-axis. The (110) surface was rotated by 55$^\circ$ relative to the (112) surface, and a small expansion ($\approx 1\%$) is applied along the $x$-direction to accommodate the non-coherent nature of the interface. A plane-wave cutoff energy of 400 eV was used in all calculations, and structural relaxations continued until the residual Hellmann–Feynman forces were below 0.03 eV Å$^{-1}$. For the bulk system, a 3 × 3 × 3 Monkhorst-Pack \textit{k}-point mesh was used to ensure adequate Brillouin zone sampling. For the larger W$\langle110\rangle$/W$\langle112\rangle$ interface cell, only the Gamma point was selected, which has been shown to provide sufficient accuracy in a previous work \cite{FERNANDEZPELLO2022153481}. Following these relaxations, the resulting equilibrium lattice constant for W was found to be 3.172 Å, in good agreement with the experimental value of 3.165 Å \cite{james1992macmillan}.

\subsection{MD simulations}
\label{subsec:Molecular_dynamics_simulations}

All MD simulations were performed using the Large-scale Atomic/Molecular Massively Parallel Simulator (LAMMPS)~\cite{plimpton1995} patched with the DeePMD-kit \cite{WANG2018178}. Alongside the DeePMD model for W that we developed, two widely used empirical IPs were employed for comparison: the EAM potential developed by Marinica \textit{et al.}~\cite{marinica2013}, commonly applied to describe tungsten–tungsten interactions in irradiation damage studies, and the Tersoff potential parameterized by Juslin and Wirth~\cite{juslin2005}, 
originally developed for covalent systems which incorporate a bond-order framework 
that captures short-range angular interactions and enables the modeling of 
complex defect structures in W. For system minimization, periodic boundary conditions (PBC) were imposed in all directions, effectively simulating an infinite crystal. A 0.5\,fs timestep was used to ensure numerical stability under defect introduction, and a conjugate gradient (CG) relaxation with zero energy and force tolerances was employed, allowing up to 10,000 steps if convergence was not reached earlier. An anisotropic relaxation scheme limited the maximum fractional deformation of the simulation box to 0.001 per minimization step.

To systematically simulate the defect structures commonly observed under irradiation conditions, the Creation-Relaxation Algorithm (CRA) \cite{PhysRevMaterials.4.023605} was used. This method efficiently generates atomic-scale microstructures by mimicking irradiation-induced defects while preserving the physical accuracy of the defect representation. In each CRA cycle, one atom was removed (creating a vacancy) and inserted at a random position at least 1.3\,\AA\ away from other atoms. The system was subsequently minimized via the CG method to achieve a near-zero stress state, effectively emulating the post-cascade defect distribution observed in irradiated materials. To ensure statistical robustness, three independent simulations were performed for each condition, with each simulation using a different random seed for defect insertion. After defect introduction, an 8\,ps run in the NVT ensemble was carried out under anisotropic pressure control (fixing $x$ and $y$, relaxing $z$), followed by an additional 100\,ps in the NVE ensemble to capture the time evolution of defect recombination and dynamics. This setup provides a comprehensive view of how GB and temperature affect defect behavior in W. The same simulation cell was used for DFT and MD point defect formation and binding energy calculations.

\subsection{FaVAD}
\label{subsec:FaVAD}

The analysis of damage in materials starts by computing the DVs 
for all the atoms in the sample. To do so, the software workflow 
 FaVAD  
\cite{VONTOUSSAINT2021107816,DOMINGUEZGUTIERREZ2020100724} was employed.
The calculation of the $i$-th atom DV, $ \vec{\xi}_{\ i}$, was 
done by considering the atom 
density $\rho_{{} i}(\vec r)$ around the $i$-th atom. 
It is expressed in terms of a sum of truncated Gaussian density 
functions as follows,

\begin{eqnarray}
\rho_{{} i}(\vec r) & = & \sum^{\textrm{neigh.}}_{j} \exp 
\left( -\frac{|\vec r-\vec r_{ij}|^2}{2 \sigma^2_{\textrm{atom}}} \right) 
f_{\textrm{cut}} \left( |\vec r_{ij}| \right)  \label{eq:Eq1} \\
     & = & \sum_{nlm}^{NLM} c^{(i)}_{nlm}g_n(r)Y_{lm}\left(\hat r\right),
\label{eq:Eq2}
\end{eqnarray}
where $\vec r_{ij}$ is the internuclear distance difference vector
between the atoms $i$ and $j$; $\sigma^2_{\textrm{atom}}$ is associated 
to the broadening of the atomic position and 
$f_{\textrm{cut}} \left( |\vec r_{ij}| \right)$ 
is a smooth cutoff function which 
removes atoms $j$ with a 
distance $|\vec r_{ij}|$ exceeding $r^{c}_{ij}$ from the 
local atomic density computation around the $i$-th atom. Here, \( r^{c}_{ij} \) denotes the cutoff radius used in the calculation of the local atomic density around the \( i \)-th atom.
$Y_{lm}(\hat r)$ are the spherical harmonics and $\hat r$ is a 
unit vector in the $\vec r$ direction \cite{VONTOUSSAINT2021107816}.
Finally, the expansion coefficients were calculated by an 
orthonormal set of radial functions
$g_n(r)$ as $c^{(i)}_{nlm} = \langle g_n Y_{lm} | \rho_i \rangle$ 
\cite{DOMINGUEZGUTIERREZ2020100724}. 
The sum over the order $m$ of the squared modulus of the coefficients 
$c_{nlm}$
is invariant under rotations around the central atom that 
defines the DV as a fingerprint of the particular local atomic
environment as 

\begin{equation}
\vec{\xi}_i^{\ k} = \left\{ \sum_m
\left(c_{nlm}^i \right)^* c_{n'lm}^i \right\}_{\ n,n',l},
\label{eq:Eq3}
\end{equation}
where $c^{*}_{nlm}$ denotes the complex conjugate of $c_{nlm}$.
Here each component $k$ of the vector $\vec{\xi}$ corresponds to 
one of the index 
triplets $\{n,n',l\}$.
The normalized DVs $\vec{q}_{i} = \vec{\xi}_{i}/|\vec{\xi}_{i}|$ were calculated with the FaVAD code \cite{VONTOUSSAINT2021107816} 
within the multi-body 
descriptor framework called SOAP \cite{Bartok2013}, as already mentioned in the text, with the 
QUantum mechanics and Interatomic Potentials (QUIP) package \cite{Csanyi2007-py}, augmented 
by the Gaussian Approximation Potential (GAP) package 
\cite{Bartok2010}.

The reference set of DVs that were used to analyze the damaged material
was obtained by 
defining a simulation cell with a thermalized and defect--free sample
(i.e., a single crystal) 
of $N$ atoms \cite{VONTOUSSAINT2021107816}. The sample was thermalized at 
room temperature (300 K) 
to take into account distortions in the atomic lattice positions 
due to thermal motions.
Thus, the mean of the reference DVs
$\vec{v}\left(T\right)=\frac{1}{N}\sum_{i=1}^{N}\vec{q}_{i}\left(T\right)$ 
together
with the associated covariance matrix $\Sigma$ of the 
descriptor--vector components
lead to estimate the weighted distance difference, $d^M (T)$, 
between two atomic environments
as \cite{VONTOUSSAINT2021107816,DOMINGUEZGUTIERREZ202238}
\begin{equation}
    d_i^M (T) =
    \sqrt{ \left( \vec{q}_{i} - \vec{v}\left(T\right) \right)^{\textrm{T}} 
\Sigma^{-1} (T) \left( \vec{q}_{i} - \vec{v}\left(T\right) \right)},
\label{eq:maha}
\end{equation} 
where the $i$-index runs for each atom in the material samples. All the irradiated samples were analyzed using the FaVAD toolkit. It is worth noting that this method is based on measuring the distances between the DVs of the atoms in our configuration and a set of reference DVs. In particular, the difference between the local environments of the $i$-th and $j$-th atoms is obtained by calculating the distance between their corresponding DVs, $d = d\left(\vec{q}_i, \vec{q}_j\right)$. Although several distance measures are conceivable, in our particular case, the Euclidean distance was used because the inverse of the estimated covariance matrix was ill-conditioned. Furthermore, as shown in previous studies, there is little advantage in using the Mahalanobis distance \cite{VONTOUSSAINT2021107816}. The definition of the distance difference, $d_M(T)$, between two local atomic environments leads to a probabilistic interpretation of the obtained results. 

\section*{Data Availability}

The datasets generated and analyzed during the current study are available from the corresponding author upon reasonable request.



\section*{Acknowledgments}

This research was funded by the Spanish Ministry of Science and
Innovation, through the predoctoral contract PRE2020-096178, by the project RADIAFUS V, grant PID2019-105325RB-C32, Ministerio de Ciencia, Innovación y Universidades, by the project PID2023-149089OB-I00 (BETMASFUS) and by the EUROFUSION grant No. ENR-IFE.01. R.I. acknowledges funding from the EIT-RM/EU project ExpSkills-REM, grant number UE-22-EXPSKILLS-21104, the AEI Project NEXPECH-2, grant number MCINN-24-PCI2024-153437 and the Agencia de Ciencia, Competitividad Empresarial e Innovación Asturiana (Sekuens) project MAGNES, SEK-25-GRU-GIC-24-113. F.J.D.G. acknowledges support from the European Union Horizon 2020 research and innovation program under grant agreement no. 857470 and from the  European Regional Development Fund via the Foundation for Polish 
Science International Research Agenda PLUS program grant  No. MAB PLUS/2018/8. P.M.P. acknowledges funding from the Marie Skłodowska-Curie Cofund Programme of the European Commission project H2020-MSCA-COFUND-2020-101034228-WOLFRAM2. P.M.P. acknowledges funding from the Marie Skłodowska-Curie Cofund Programme of the European Commission, project H2020-MSCA-COFUND-2020-101034228-WOLFRAM2. We also acknowledge support provided by the Eu-MACE and EuMINe COST Actions CA22123 and CA22143, respectively. 
Simulations reported here were substantially performed using the Princeton Research Computing resources at Princeton University.

\section*{Author contributions}

J.S.-R., R.I., R.G.-A. and P.M.P. designed the research. J.S.-R. performed the simulations. J.S.-R., P.M.P. and F.J.D.-G. analyzed the data. J.S.-R., P.M.P. and F.J.D.-G. wrote the paper. R.I. and R.G.-A. supervised the project. All authors discussed the results and commented on the manuscript.

\section*{Competing interests}
The authors declare no competing interests.








\end{document}